\documentclass[numberedreferences]{kluwer}
\usepackage{graphicx}

\def\kms{km~s$^{-1}$}
\begin{document}
\begin{article}
\begin{opening}
\title{The tiny-scale atomic structure: gas cloudlets or scintillation
phenomenon ?}

\author{S.
\surname{Stanimirovi\'{c}}\email{sstanimi@astro.berkeley.edu}}
\institute{Radio Astronomy Lab, UC Berkeley, 601 Campbell Hall,
Berkeley, CA 94720}
\author{J. M. \surname{Weisberg}}
\institute{Department of Physics and Astronomy, Carleton College,
Northfield,
MN 55057}
\author{A. \surname{Hedden}}
\author{K. \surname{Devine}}
\author{T. \surname{Greent}}
\institute{Department of Physics and Astronomy, Carleton College,
Northfield,
MN 55057}
\author{S. B. \surname{Anderson}}
\institute{Department of Astronomy, MS 18-34, California Institute of
Technology, Pasadena, CA 91125}


\runningtitle{The TSAS: gas cloudlets or scintillation phenomenon ?}
\runningauthor{Stanimirovi\'{c} et al.}

\begin{abstract}
We present here preliminary results from the recent multi-epoch HI
absorption measurements toward several pulsars using the Arecibo
telescope. We do not find significant variations in optical depth
profiles over periods of 4 months and 9 years. 
The upper limits on the optical depth variations in directions of B0823+26
and B1133+16 are similar on scales of 10--20, 350 and 500 AU.
The large number of non detections of the tiny scale atomic structure
suggests that the AU-sized structure is not ubiquitous in the 
interstellar medium.
\end{abstract}


\end{opening}

\section{Introduction}
For many years there have been theoretical and observational support for
the existence of structure in the interstellar medium (ISM) on scales
from $\sim$ 1 pc to 1 kpc (as summarized by Dickey \& Lockman 1990),
while structure on
scales smaller than 1 pc was not believed to have a significant role.
As shown in Heiles(2000) it is
expected that structure on very small scales ($<1$ pc) is not prominent
in the ISM. Indeed, if we assume the standard thermal pressure for the
cold neutral medium (CNM) $P_{\rm th} \sim 2250$ cm$^{-3}$ K
(Jenkins \& Tripp 2001), and the mean
temperature for the CNM of $\sim$70 K (Heiles \& Troland 2003),
then the expected volume density for the CNM clouds is about 30
cm$^{-3}$.
With the mean measured column density of 1--5$\times10^{20}$
cm$^{-2}$ (Heiles \& Troland 2003), the typical expected scale lenght
for the CNM features is $\sim1$ pc.
It was therefore quite surprising to frequently find structure on 
AU scales in many different directions in the ISM.

The AU-sized structure is observationally probed using spatial and/or
temporal variations in absorption profiles against background sources.
There are several varieties of this observational technique.
For the atomic medium, the following three techniques are commonly used.
(1) Spatial variation of HI absorption line profiles across the
extended background source (Dieter, Welch, \& Romney 1976;
Diamond et al. 1989; Davis, Diamond, \& Goss 1996; Faison \& Goss 2001) 
typically probes scales of a few tens to
a few hundreds of AUs.
(2) Comparison of optical interstellar absorption lines against binary
stars and globular clusters (Meyer \& Blades 1996; Lauroesch et al. 1998) 
probes larger scales of 10$^{2}$--10$^{6}$ AU.
(3) Time variability of HI absorption profiles against pulsars
(Deshpande et al. 1992; Frail et al. 1994; Johnston et al. 2003)
probes scales of tens to hundreds of AUs.
In the case of the molecular medium, the fine-scale structure is usually
observed through
time and/or spatial variability of molecular line absorption profiles
against compact extragalactic sources (Marscher, Moore, \& Bania 1993;
Moore \& Marscher 1995) 
and typically probes scales of tens of AUs.
In this paper we concentrate on the AU-sized structure in the cold
atomic medium, often referred to as the tiny-scale atomic structure
(TSAS).

The direct quantities provided by observations of the TSAS
are optical depth variation ($\Delta \tau$) and a particular time
or spatial baseline, which can be translated into transverse
size ($L_{\perp}$). Two additional constraints are
the frequency of detection and the variability of the line shape.
As the AU-sized structure was very frequently detected it was thought
that it is more likely to be a general property of the ISM than the effect
of some local phenomena. As noted by Heiles (1997), it is often that the
line depth was found to vary rather than the line shape, suggesting that
the TSAS is kinematically related to the CNM.

The straight forward and traditional way of interpreting
above mentioned observations is that optical depth variations are
due to a blob moving in and out of the line-of-sight, whose transverse
dimension is equal to $L_{\perp}$. Assuming a simple spherical geometry,
the HI volume density of these blobs can be estimated and is typically
of order of $10^{4}$ cm$^{-3}$, which is very dense.
The inferred thermal pressure is then of the order of
$10^{6}$ cm$^{-3}$ K, much higher than the hydrostatic equilibrium
pressure of the ISM or the standard thermal pressure of the CNM. This
has been known for a long time and has caused much controversy. It is
expected that such over-dense and over-pressured
features should dissipate on a time scale of about 100 yrs and 
therefore not be common in the ISM. A variety of other troubling
questions concerning the AU-sized structure includes the following.
How are AU-sized features formed and maintained in the ISM?
What is the fraction of the ISM occupied by these features?
Is this is a new population of interstellar clouds? And is there only
one
such population or many? How do these features relate to the continuous
hierarchy of structure observed on larger spatial scales?

In order to solve this long-standing puzzle several alternative
explanations were proposed.
Heiles (1997) suggested that
TSAS features are actually curved filaments and/or sheets that happen to
be aligned along our line-of-sight. The ratio of the line-of-sight to
the plane-of-the-sky length of 4--10 is required to bring the TSAS HI
volume density to modest, acceptable,  values.
Deshpande (2000) suggested that TSAS blobs
correspond to the tail of a hierarchical structure organization that
exists on larger scales. They pointed out that TSAS observations were
misinterpreted by associating measured $L_{\perp}$ with the longitudinal
dimension of TSAS clouds, leading to extraordinary high volume
densities. 
Gwinn (2001) proposed that optical depth fluctuations
seen in multi-epoch pulsar observations are actually a scintillation
phenomenon combined with the velocity gradient across the absorbing HI
of order of 0.05--0.3 \kms AU$^{-1}$.
Different explanations predict a different level
of optical depth variations at a particular scale size. For example,
Deshpande (2000) expects that optical depth variations would increase
with the size of structure, while Gwinn (2001) predicts maximum
variations on very small spatial scales probed by the interstellar 
scintillation.
All suggested explanations, however, call for more observational data.

Motivated by the recent theoretical efforts in understanding the nature and
origin of the TSAS we have undertaken new multi-epoch observations of HI
absorption against a set of bright pulsars.  We decided to observe the same
sources as Frail et al. (1994) in order to enhance the number of available
time baselines for comparison.  
This paper summarizes preliminary results from this project.

\section{Observations and Data Processing}

We have used the Arecibo telescope\footnote{The Arecibo Observatory
is part of the National Astronomy
and Ionosphere Center, operated by Cornell University under a
cooperative agreement with the National Science Foundation.}
to obtain new multi-epoch HI absorption measurements against six
pulsars previously studied by Frail et. al. (1994).
For detail observing and data processing description see
Stanimirovic et al. (2003).
We had four observing sessions: August 2000, December 2000,
September 2001 and November 2001, measuring HI absorption profiles over
time
intervals from less than a day to 1.25 years.
The Caltech Baseband Recorder was used as a fast-sampling backend, with
a total bandwidth of 10 MHz, recording the raw voltage data every 100
ns. The first stage of data reduction was performed at the Caltech's
Center
for Advance Computation and Research.
The `pulsar-on' and `pulsar-off' spectra were accumulated by finding the
pulsar pulse and extracting
spectra during the pulse and between pulses, respectively. The pulsar
absorption spectrum is created by generating the `pulsar-on' --
`pulsar-off' for each scan, doing frequency switching to flat the
baseline,
and accumulating all such spectra with a weight proportional to $T_{\rm
PSR}$, the brightness temperature of the pulsar.
Final absorption and emission spectra have velocity resolution of 0.5
\kms.

\section{Results}

\begin{figure}
\centerline{\includegraphics[width=3.6in]{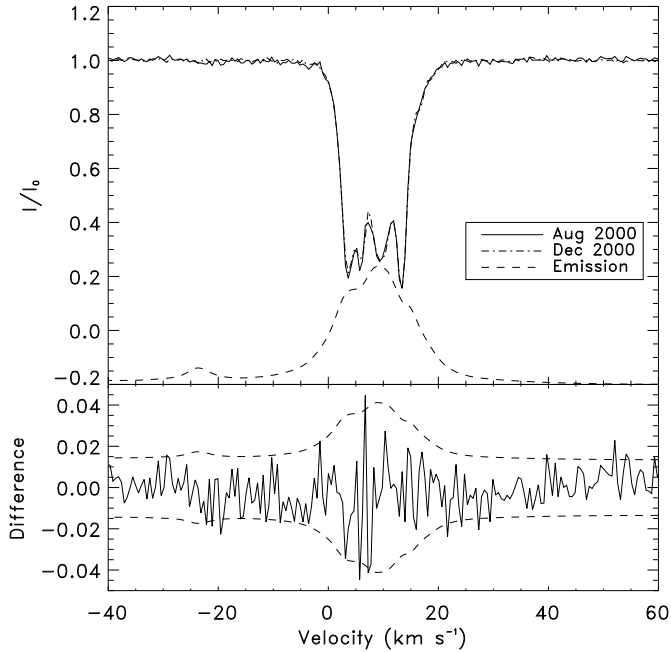}
}
\caption{{\it Top:} HI absorption spectra against B2016+28 observed in
August and December 2000. The HI emission spectrum, obtained in the 
same direction, has been multiplied by 0.0005. 
{\it Bottom:} Difference between the two
absorption spectra. Contours show $\pm$2-$\sigma$ levels where the HI
brightness temperature has been taken into account.}
\end{figure}

\begin{figure}
\centerline{\includegraphics[width=3.6in]{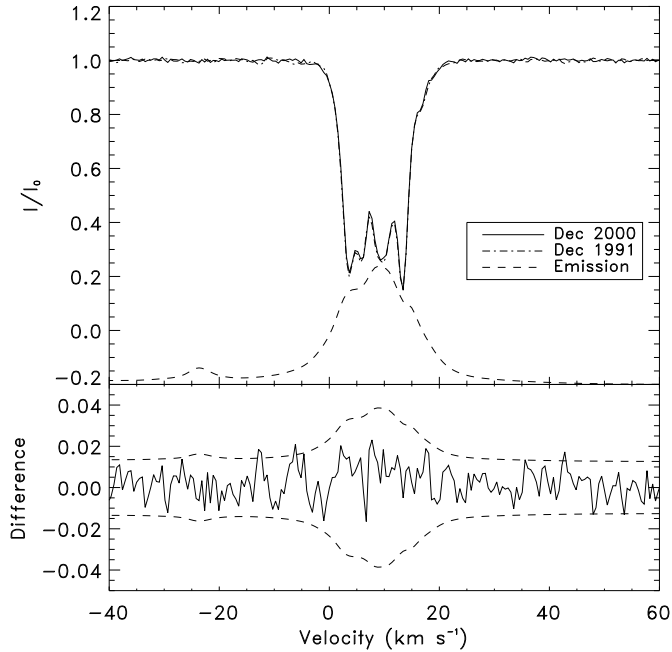}
}
\caption{Comparison of HI absorption profiles against B2016+28 obtained
in December 1991 and December 2000. Details are the same as in Figure 1.}
\end{figure}

We present here the results for three pulsars: B0823+26, B1133+16, and
B2016+28, from two observing sessions in August 2000 and December 2000. In
addition, we have used the third session obtained by Frail et al. (1994) to
provide a long term comparison. Fig. 1 shows absorption and emission
spectra for PSR B2016+28 from Aug 2000 and Dec 2000, while Fig. 2 compares
data from Aug 2000 and Dec 1991. The top panel in each figure shows
absorption spectra obtained at two different epochs, and a scaled emission
spectrum to get a feeling about how noise in the absorption spectra varies
with velocity.  The bottom panel in each plot shows a difference of
absorption profiles at two epochs with overlaid $\pm$ 2-$\sigma$
significance envelopes.  To estimate the rms noise we have taken into
account contribution from the sky background and, very importantly,
contribution from the HI brightness temperature which increases the rms
noise on line significantly (for example, by a factor of 3 in the case of
B2016+28).

For all three pulsars we compared HI absorption spectra over a time span of
4 months (Aug 2000 to Dec 2000) and 9 years (Dec 1991 to Dec 2000). Except
for the case of B2016+28 between Aug 2000 and Dec 2000, in all other cases
we {\it do not} find significant change in the absorption spectra.
B2016+28 shows a marginal, 2.5-$\sigma$, change in absorption spectra
obtained with a 4 months difference.  Transverse velocity of this object is
39 \kms, meaning that during the period of 4 months B2016+28 has
traveled over a distance of 3 AU. This marginal detection could be caused
by structure on scales of 3 AU.

Other comparisons do not show change in absorption spectra on scales of
about 10-20, 70
and 350-500 AU at $\Delta\tau$ level down to about 0.02. Information on
individual objects and an upper limit on $\Delta\tau$ are given in Table
1 (note that these are 2-$\sigma$ limits on the HI line).
Frail et al. (1994) found variations of 0.07 for B0823+16 and 0.04 for
B1133+16 over a period of 13 months. In the case of B2016+28
they noticed even larger optical depth variations of almost 1 over  7
and 20 months periods.
Our sensitivity is better than typical variations seen in Frail et
al. (1994) and our non detections are therefore a significant result.
As already pointed out by Johnston et al. (2003) there is a serious
concern that the Frail et al.'s data have been over interpreted.
For example, in the case of B2016+28 the noise level on line is at least
three times higher than the noise level off line, due to the HI
brightness
temperature, while Fig. 2 of Frail et al. (1994) claims an increase of
no more than 1.5.

The upper limits on optical depth fluctuations set by B0823+26 and B1133+16
at scales of 10--20 and 350--500 AU are strikingly low.  $\Delta\tau$ of
0.03 corresponds to $3\times10^{18}$ cm$^{-2}$ for the column density
fluctuations of the CNM.  In addition, these upper limits are almost the
same for spatial scales that are more than an order of magnitude different!
Deshpande (2000) predicts opacity variations as an extension of HI opacity
irregularities observed on larger scales using a single power law spectrum.
Using the power spectrum of opacity distribution in the direction of Cas A
and extrapolating down to AU scales, they predict  $\Delta\tau \sim 0.03$
at scales of about 10 AU, while at 50--100 AU $\Delta\tau \sim$0.2--0.4 is
expected.  This model generally expects that $\Delta\tau$ increases with
spatial scales.  While results for B0823+26 and B1133+16 at scales of
10--20 AU are consistent with this picture, our upper limits for $\Delta
\tau$ at scales of 350 and 500 AU for the same pulsars are significantly
lower from what is expected. However, this is still a very small number
statistics and we will be able to provide stronger constraints on
particular theoretical models in the near future. In addition, it is
important to mention that the power spectrum slope derived for the
direction toward Cas A may not be applicable to directions sampled by
pulsars in this study.

This large number of TSAS non detections, together with recent results by
Johnston et al. (2003), is disturbing and unexpected!  While previous
observations were frequently detecting TSAS, in six comparisons presented
here, and in many comparisons in Johnston et al. (2003), there is only one
marginal detection for the case of B2016+28 and one detection for B1641-45
(Johnston et al. 2003) over a period of almost twenty years.  This rises
the question of the existence of TSAS as traced at least by multi-epoch
pulsars observations. Contrary to the previous belief that the TSAS is
ubiquitous in the ISM, our results indicate that TSAS may be more rare and
could be related to some kind of a local phenomenon.

\section{Conclusions}

\begin{table}
\caption{Transverse scales and maximum $\Delta \tau$ probed by the three
pulsars. Note that Max. $\Delta \tau$ is a 2-$\sigma$ limit estimated on the HI
line.  }
\label{table1}
\begin{tabular}{lll}
\hline
PSR      & $L_{\perp}$ (AU) & Max. $\Delta \tau$ \\
\hline
B0823+26 & 10        & $<0.04$ \\
         & 350       & $<0.04$ \\
B1133+16 & 20        & $<0.03$    \\
         & 500       & $<0.02$    \\
B2016+28 & 3         & 0.18    \\
         & 70        & $<0.2$    \\
\hline
\end{tabular}
\end{table}

We have compared multi-epoch HI absorption observations toward \\B0823+26,
B1133+16 and B2016+28 over time periods of 4 months and 9 years. Except
for a marginal change in the case of B2016+28 over a 4-month period, we
do not find significant changes in absorption spectra.
This is very different from previous observations by Frail et al. (1994)
who saw significant optical depth variations for the same pulsars over
time periods of 7, 13 and 20 months. 
We have placed upper limits on the optical depth variations of 0.02--0.04 in
directions of B0823+26 and B1133+16. In addition, the upper limit on
optical depth variations appears the same for spatial scales of 10--20,
350 and 500 AU, traced by these objects.  
A large number of non detections of the TSAS presented here, together
with recent results by Johnston et al. (2003), suggests that the TSAS is
not ubiquitous in the ISM.

\begin{acknowledgements}
We are grateful to Caltech's Center for Advance Computation and Research
for the use of their facilities for data storage and processing.
S. S. acknowledges support by NSF grants  AST-0097417 and AST-9981308.
\end{acknowledgements}

\end{article}

\begin{thebibliography}{00}


\bibitem{Davis96}
Davis, R.~J., P.~J. Diamond, and W.~M. Goss: 1996, `{MERLIN and EVN 
observations of Small-Scale Structure in the Interstellar HI}'.
\newblock {\it Monthly Notices of the Royal Astronomical Society} 
{\bf 283}, 1105.

\bibitem{Deshpande00a}
Deshpande, A.~A.: 2000, `{The small-scale structure in interstellar HI: 
a resolvable puzzle}'.
\newblock {\it Monthly Notices of the Royal Astronomical Society} 
{\bf 317}, 199D.

\bibitem{Deshpande92}
{Deshpande}, A.~A., P.~M. {McCulloch}, V. {Radhakrishnan}, and K.~R.
  {Anantharamaiah}: 1992, `{Pulsar observations and small-scale structure in
  interstellar H I}'.
\newblock {\it Monthly Notices of the Royal Astronomical Society} 
{\bf 258}, 19P--21P.

\bibitem{Diamond89}
Diamond, P.~J., W.~M. Goss, J.~D. Romney, R.~S. Booth, P.~N.~M. Karbela, and U.
  Mebold: 1989, `{The structure of the interstellar medium at the 25 AU scale}'.
\newblock {\it Astrophysical Journal} {\bf 347}, 302.

\bibitem{Dickey90}
Dickey, J.~M. and F.~J. Lockman: 1990, `{H I in the Galaxy }'.
\newblock {\it Annual review of astronomy and astrophysics} {\bf 28}, 215.

\bibitem{Dieter76}
Dieter, N.~H., W.~J. Welch, and J.~D. Romney: 1976, `{A very small 
interstellar neutral hydrogen cloud observed with VLBI techniques}'.
\newblock {\it Astrophysical Journal} {\bf 206}, L113.

\bibitem{Faison01}
{Faison}, M.~D. and W.~M. {Goss}: 2001, `{The Structure of the Cold Neutral
  Interstellar Medium on 10-100 AU Scales}'.
\newblock {\it Astronomical Journal} {\bf 121}, 2706--2722.

\bibitem{Frail94}
Frail, D.~A., J.~M. Weisberg, J.~M. Cordes, and C. Mathers: 1994, `{Probing 
the interstellar medium with pulsars on AU scales }'.
\newblock {\it Astrophysical Journal} {\bf 436}, 144.

\bibitem{Gwinn01}
{Gwinn}, C.~R.: 2001, `{Small-Scale Variations of H I Spectra from Interstellar
  Scintillation}'.
\newblock {\it Astrophysical Journal} {\bf 561}, 815--822.

\bibitem{Heiles97}
Heiles, C.: 1997, `{Tiny-Scale Atomic Structure and the Cold Neutral Medium}'.
\newblock {\it Astrophysical Journal} {\bf 481}, 193.

\bibitem{Heiles03b}
{Heiles}, C. and T.~H. {Troland}: 2003, `{The Millennium Arecibo 21 Centimeter
  Absorption-Line Survey. II. Properties of the Warm and Cold Neutral Media}'.
\newblock {\it Astrophysical Journal} {\bf 586}, 1067--1093.

\bibitem{Heiles00}
Heiles, C.~E.: 2000.
\newblock In: D.~G. Finley and W.~M. Goss (eds.): {\it Radio interferometry :
  the saga and the science}. p.~7.

\bibitem{Jenkins01}
{Jenkins}, E.~B. and T.~M. {Tripp}: 2001, `{The Distribution of Thermal
  Pressures in the Interstellar Medium from a Survey of C I Fine-Structure
  Excitation}'.
\newblock {\it Astrophysical Journal Supplement Series} {\bf 137}, 297--340.

\bibitem{Johnston03}
{Johnston}, S., B. {Koribalski}, W. {Wilson}, and M. {Walker}: 2003,
  `{Multi-epoch HI line measurements of four southern pulsars}'.
\newblock {\it Monthly Notices of the Royal Astronomical Society} 
{\bf 341}, 941--947.

\bibitem{Lauroesch98}
{Lauroesch}, J.~T., D.~M. {Meyer}, J.~K. {Watson}, and J.~C. {Blades}: 1998,
  `{The Physical Characteristics of the Small-Scale Interstellar Structure
  toward MU Crucis}'.
\newblock {\it Astrophysical Journal} {\bf 507}, L89--L92.

\bibitem{Marscher93}
Marscher, A.~P., E.~M. Moore, and T.~M. Bania: 1993, `{Detection of 
AU-Scale Structure in Molecular Clouds}'.
\newblock {\it Astrophysical Journal} {\bf 419}, L101.

\bibitem{Meyer96}
{Meyer}, D.~M. and J.~C. {Blades}: 1996, `{Small-Scale Interstellar Medium
  Structure: The Remarkable Sight Line toward MU Crucis}'.
\newblock {\it Astrophysical Journal} {\bf 464}, L179.

\bibitem{Moore95}
Moore, E.~M. and A.~P. Marscher: 1995, `{Observational Probes of 
the Small-Scale Structure of Molecular Clouds}'.
\newblock {\it Astrophysical Journal} {\bf 452}, 671.

\bibitem{Stanimirovic03}
{Stanimirovi{\' c}}, S., J. {Weisberg}, J.~M. {Dickey}, A. {de la Fuente}, K.
  {Devine}, A. {Hedden}, and S.~B. {Anderson}: 2003, `{Detection of OH
Absorption Against PSR B1849+00}'.
\newblock {\it Astrophysical Journal} {\bf 592}.
\newblock in press.


\end{thebibliography}
\end{document}